\documentclass[12pt,twoside]{article}
\usepackage{mathpazo} 
\usepackage[mathscr]{eucal}
\usepackage{amsmath,amsfonts,amssymb,amsthm}
\usepackage[matrix,arrow]{xy}
\usepackage{times}
\usepackage{pdfsync}
\usepackage{graphicx}
\usepackage{epstopdf}
\usepackage{dcolumn}
\usepackage{hyperref}
\usepackage{color}

\def\d_Vphi{\text{d}_V\hspace{-0.06em}\phi}
\def\d_Vphibar{\text{d}_V\hspace{-0.06em}\bar\phi}
\def\d_Vxi{\text{d}_V\hspace{-0.06em}\xi}

\def\be{\begin{eqnarray}}
\def\ee{\end{eqnarray}}
\def\beann{\begin{eqnarray*}}
\def\eeann{\end{eqnarray*}}
\def\beq{\begin{equation}}
\def\eeq{\end{equation}}
\def\ba{\begin{array}}
\def\ea{\end{array}}
\def\ben{\begin{enumerate}}
\def\een{\end{enumerate}}
\def\bea{\begin{eqnarray}}
\def\eea{\end{eqnarray}}

\def\5{\bar }
\def\6{\partial }
\def\7{\hat }
\def\4{\tilde }
\advance\textheight\topskip
\renewcommand{\tilde}{\widetilde}
\renewcommand{\hat}{\widehat}

\newcommand{\bref}[1]{\textbf{\ref{#1}}}

\newcommand{\dd}{\partial}
\renewcommand{\d}{\partial}

\renewcommand{\geq}{\,{\geqslant}\,}

\newcommand{\binner}[2]{%
  {\langle}\kern-4.15pt{\langle}#1{,}\,#2{\rangle}\kern-4.15pt{\rangle}}

\newcommand{\half}{\frac{1}{2}}

\newcommand{\ffrac}[2]{\raisebox{.5pt}%
  {\footnotesize$\displaystyle\frac{#1}{#2}$}\kern1pt}

\newcommand{\ddl}[2]{\ffrac{\dd #1}{\dd #2}}

\def\cF{\mathcal{F}}

\def\cH{\mathcal{H}}

\def\cJ{\mathcal{J}}

\numberwithin{equation}{section} \makeatletter
\@addtoreset{equation}{section}

\DeclareFontFamily{OT1}{rsfs}{} \DeclareFontShape{OT1}{rsfs}{m}{n}{
<-7> rsfs5 <7-10> rsfs7 <10-> rsfs10}{}
\DeclareMathAlphabet{\mycal}{OT1}{rsfs}{m}{n}
\def\scri{{\mycal I}}%

\begin{document}

\def\mytitle{Entropy of three-dimensional asymptotically flat
  cosmological solutions}

\pagestyle{myheadings} \markboth{\textsc{\small Glenn Barnich}}{%
  \textsc{\small Entropy of 3d asymptotically flat cosmological
    solutions}} \addtolength{\headsep}{4pt}

\begin{flushright}\small
ULB-TH/12-14\end{flushright}

\begin{centering}

  \vspace{1cm}

  \textbf{\Large{\mytitle}}

\vspace{1cm}

  \vspace{1.5cm}

  {\large Glenn Barnich$^{a}$}

\vspace{.5cm}

\begin{minipage}{.9\textwidth}\small \it \begin{center}
   Physique Th\'eorique et Math\'ematique\\ Universit\'e Libre de
   Bruxelles\\ and \\ International Solvay Institutes \\ Campus
   Plaine C.P. 231, B-1050 Bruxelles, Belgium \end{center}
\end{minipage}

\end{centering}

\vspace{1cm}

\begin{center}
  \begin{minipage}{.9\textwidth}
    \textsc{Abstract}. The thermodynamics of three-dimensional
    asymptotically flat cosmological solutions that play the same role
    than the BTZ black holes in the anti-de Sitter case is derived and
    explained from holographic properties of flat space. It is shown
    to coincide with the flat-space limit of the thermodynamics of the
    inner black hole horizon on the one hand and the semi-classical
    approximation to the gravitational partition function associated
    to the entropy of the outer horizon on the other. This leads to
    the insight that it is the Massieu function that is universal in
    the sense that it can be computed at either horizon.
  \end{minipage}
\end{center}

\vfill

\noindent
\mbox{}
\raisebox{-3\baselineskip}{%
  \parbox{\textwidth}{\mbox{}\hrulefill\\[-4pt]}}
{\scriptsize$^a$Research Director of the Fund for Scientific
  Research-FNRS Belgium. E-mail: gbarnich@ulb.ac.be}

\thispagestyle{empty}
\newpage


\section{Introduction}
\label{sec:introduction}

In order to test holographic ideas in gravitational theories
\cite{tHooft:1993gx,Susskind:1994vu} and go beyond the context of the
${\rm AdS/CFT}$ correspondence \cite{Maldacena:1998re}, it seems
useful to first try to extend the ${\rm AdS}_3$ results to the flat
case. Indeed, for the former case, there is complete control on symmetries,
charges and central extensions \cite{Brown:1986nw}, on solution space
\cite{1999AIPC..484..147B,Skenderis:1999nb}, including black holes
\cite{Banados:1992wn,Banados:1993gq} and a compelling conformal field
theory interpretation \cite{Strominger:1998eq}.

Asymptotically flat gravity in three dimensions at null infinity
\cite{witten:98xx} is arguably as simple and interesting a model:
again, there is complete control on symmetries \cite{Ashtekar:1996cd},
charges, central extensions \cite{Barnich:2006avcorr}, solution space
with a conformal field theory interpretation \cite{Barnich:2010eb},
and the precise relation to the ${\rm AdS}_3$ case
\cite{Barnich:2012aw}. In particular, the flat-space limit of the BTZ
black holes are simple cosmological solutions.

The purpose of this paper is to try push this project to a level of
understanding similar to that achieved in the ${\rm AdS}_3$ case by
deriving the thermodynamics of the Cauchy horizon of the cosmological
solutions and providing a holographic derivation of their entropy in
the grand canonical ensemble through an appropriate Cardy-like
formula.

We then point out that this thermodynamics is the flat-space limit of
the one of the inner BTZ horizon. More generally, the relevance of
inner horizon thermodynamics in this and related contexts has been
stressed for instance in
\cite{Cvetic:1997uw,Park:2006pb,Cvetic:2010mn,Castro:2012av,%
Detournay:2012ug,Chen:2012mh}.
The semi-classical approximation to the logarithm of the partition
function of the ${\rm AdS}_3$ case is a Massieu function for the
entropy of the outer horizon and has a good flat-space limit that
coincides with the one directly computed in the flat case.

As a by-product and a case in point, it is then readily seen that the
Massieu function is universal in the sense that it can be computed
from the entropy at either BTZ horizon.

\section{Thermodynamics of cosmological solutions}
\label{sec:cosm-solut}

In BMS form, the cosmological solutions are explicitly given by 
\begin{equation}
ds^2=8GM du^2-2du
  dr+8GJ du d\phi+r^2d\phi^2\label{eq:Squad},
\end{equation}
with $M>0$, while their  ADM form is
\begin{equation}
  \label{eq:26}
\begin{gathered}
  ds^2=-N^2dt^2+N^{-2}dr^2+r^2(d\varphi+N^\varphi dt)^2,\\ 
N^2=-8MG +\frac{16 G^2 J^2}{r^2},\quad
N^\varphi=\frac{4GJ}{r^2},
\end{gathered}
\end{equation}
where, again, $t=u+f(r)$, $\varphi=\phi+g(r)$, with $f^\prime= N^{-2}$,
$g^\prime=-N^\varphi f^\prime$. In the discussions below, we explicitly
assume $J\neq 0$ most of the time. 
In ADM coordinates, it follows that the null hypersurface
\begin{equation}
r_C=\sqrt{\frac{2G J^2}{M}}\label{eq:87},
\end{equation}
is special, but in BMS coordinates, which now play the role of
outgoing Eddington-Finkelstein coordinates, this hypersurface is
regular.

As discussed in \cite{Barnich:2012aw}, let $\alpha=\sqrt{8GM}$ and
consider the coordinate changes\footnote{Note that there is a sign
  mistake in equation (61) of this reference, the first term on the
  right hand side should read $(\alpha dt+\frac{4GJ}{\alpha}
  d\varphi)^2$}, $X=\varphi-\Omega_C t$, so that
$(\varphi,X)\sim(\varphi+2\pi,X+2\pi)$,
$T^2=\frac{1}{\alpha^2}(r^2-r^2_C)$ for $r> r_C$, and $\bar
r^2=-\frac{1}{\alpha^2}(r^2-r^2_C)$ for $r<r_C$. The metric then
becomes
\begin{eqnarray}
  \label{eq:23}
  ds^2=\left\{\begin{array}{l} -dT^2+r^2_C dX^2 +\alpha^2
      T^2d\varphi^2, \quad r>r_H,\\
       d\bar r^2+r^2_C dX^2-\alpha^2 \bar r^2 d\varphi^2,\quad r<r_H.
\end{array}\right.
\end{eqnarray}
In the outer region, it thus describes a cosmology with spatial
section a torus with radii $r_C$ and $\alpha T$. Furthermore, the
curves $\varphi=\lambda= X$, $T=cte$ respectively $\bar r=cte$ are
closed geodesics that are spacelike in the outer region and time-like
in the inner region when $\bar r> \frac{r_C}{\alpha}$. It follows that
the hypersurface is a Cauchy horizon and that one may decide to cut
the space-time at $\bar r=\frac{r_C}{\alpha}$ and thus at $r=0$, as in
the BTZ case. The proof in \cite{Banados:1993gq} on the absence of
closed time-like curves using ADM coordinates can then directly be
applied to this case as well, the only difference being that the outer
region (I) of the BTZ black hole has disappeared as the outer horizon
is pushed to infinity in the flat limit.

From the point of view of identifications of Minkowski spacetime,
these geometries have been studied previously in
\cite{Ezawa:1992nk,Cornalba:2002fi,Cornalba:2003kd}. In the latter two
references for instance, their role for string cosmology and how they
arise as a suitable limit of BTZ black holes have been emphasized.

The Cauchy horizon is also a Killing horizon. The generator is
$\xi=\d_u+\Omega_C\d_\phi=\d_t+\Omega_C\d_\varphi$, while the surface
gravity is determined through $\xi^\nu D_\nu \xi^\mu=\kappa_C
\xi^\mu$, with
\begin{equation}
  \label{eq:88}
  \mu_C=\Omega_C=-\frac{2M}{J}, \quad T_C=\beta_C^{-1}=
  \frac{\kappa_C}{2\pi}=\frac{\Omega^2_C}{2\pi }
  r_C=\frac{2}{\pi}\sqrt{\frac{2GM^3}{J^2}}. 
\end{equation}
Let $\Phi_C=\beta_C\mu_C=-\frac{2\pi}{\alpha}\frac{|J|}{J}$.
In terms of the Bekenstein-Hawking entropy
\begin{equation}
  \label{eq:89}
  S_C=\frac{2\pi r_C}{4G}=\frac{\pi^2}{G\beta_C\mu_C^2}=\frac{\pi^2\beta_C}{G\Phi_C^2},
\end{equation}
the first law takes the form 
\begin{equation}
dM=-T_C dS_{C}-\Omega_C dJ\label{eq:90}.
\end{equation}

Define $\ln Z_C(\beta_C,\Phi_C)$ as the Legendre transform of the
entropy $S_C(M,J)$ that satisfies $M=-\ddl{\ln Z_C}{\beta_C}$,
$J=-\ddl{\ln Z_C}{\Phi_C}$. It is a (generalized) Massieu function for
the entropy and the semi-classical approximation of the partition
function (see e.g.~\cite{Callen} and also
\cite{Gibbons:1976ue,0264-9381-7-8-020} for considerations in
Euclidean quantum gravity).

 The unusual form of the first law forces one to use the opposite sign
 for $S_C$ in the Legendre transform as compared to the case of a
 standard first law,
 \begin{equation}
   \label{eq:14}
  \ln Z_{C}(\beta_C,\Phi_C)=- S_C-\beta_C M-\Phi_C J, \quad \ln
   Z_{C}=-\frac{\pi^2\beta_C}{2G\Phi_C^2}=-\frac{\pi^2}{2G\beta_C\mu_C^2}. 
 \end{equation}

 Note that if one takes the Hawking temperature to be negative,
 $T_C=\beta_C^{-1}=-\frac{2}{\pi}\sqrt{\frac{2GM^3}{J^2}}$, the first
 law comes with the usual sign, which implies that it is $S_C$ rather
 than $-S_C$ that is involved in the Legendre transform. The final
 expression for $\ln Z_C$ is unchanged.

 \section{Euclidean solution}
 \label{sec:euclidean-solution}

 When letting $t=-it_E$, $J=iJ_E$, $\alpha=i\alpha_E$,
 $r^2_{EC}=\frac{16 G^2 J^2_E}{\alpha^2_E}$, the corresponding Euclidean
 solution is
   \begin{equation}
     \label{eq:13}
     ds_E^2=\frac{\alpha_E^2(r^2-r^2_{EC})}{r^2}dt^2_E+\frac{r^2}{\alpha_E^2(r^2-r^2_{EC})}dr^2+
 r^2(d\varphi+\frac{4GJ_E}{r^2} dt_E)^2.
   \end{equation}
   Let $\epsilon_J$ denote the sign of $J$. The change of coordinates
 \begin{eqnarray}
 \left\{\begin{array}{l}
 R^2_E=\frac{1}{\alpha^2_E}(r^2-r^2_{EC}),\\
 \varphi_E=-\epsilon_J\alpha _E \varphi, \\
 Z_E=r_{EC} \varphi+ \Omega_{EC} r_{EC} t_E
 \end{array}\right. \iff \left\{
 \begin{array}{l} r^2=\alpha^2_ER^2_E +r^2_{EC},\\
 \varphi=-\frac{\epsilon_J}{\alpha_E}\varphi_E,\\
 t_E=\frac{\epsilon_J}{\alpha_E\Omega_{EC}}\varphi_E+\frac{1}{\Omega_{EC} r_{EC}}Z_E,
 \end{array}\right. 
 \label{eq:1}
 \end{eqnarray}
 where $\Omega_{EC}=-\frac{2M}{J_E}=i\Omega_C$
 brings the metric explicitly to the flat form 
 \begin{equation}
   \label{eq:2}
   ds_E^2=dZ^2_E+dR^2_E+R^2_E d\varphi^2_E.
 \end{equation}
 Absence of conical singularities requires $\varphi_E\sim\varphi_E+
 2\pi$ and implies $\varphi\sim \varphi+\Phi_{EC}$, $t_E\sim
 t_E+\beta_{EC}$, where
 $\Phi_{EC}=-\epsilon_J\frac{2\pi}{\alpha_E}=i\Phi_C$ and
 $\beta_{EC}=\epsilon_J\frac{2\pi }{\Omega_{EC}\alpha_E}=\beta$, in
 agreement with section \bref{sec:cosm-solut}.

 \section{3d flat space holography }
 \label{sec:asympt-flat-3d}

 \subsection{${\rm BMS}_3$ algebra and group}
 \label{sec:bms3-group}

 Let $Y=Y(\phi), T=T(\phi)$. Following
 \cite{Ashtekar:1996cd,Barnich:2010eb,Barnich:2011ct}, the
 $\mathfrak{bms}_3$ algebra can be represented in terms of vector fields in one,
 two and three dimensions, namely as 

 \begin{enumerate}

 \item the semi-direct sum of algebra of vector fields
   $y=Y\partial_\phi$ on the circle with the abelian ideal of tensor
   densities of degree $-1$, $t=Td\phi^{-1}$, where
   $[y,t]=Y\partial_\phi T+\partial_\phi Y T$,

 \item the Lie algebra of vector fields
 $\xi=(T+u\partial_\phi Y)\partial_u +Y\partial_\phi$ on
 $\scri^+=S^1\times \mathbb{R}$ with coordinates $(u,\phi)$,

 \item the algebra of vector fields describing the symmetries of
   asymptotically flat three-dimensional spacetimes at null infinity
   equipped with the Lie algebroid bracket.

 \end{enumerate}

 The basis elements
 $j_m\leftrightarrow (Y= e^{im\phi}, T=0)$ and  $p_m \leftrightarrow
 (Y=0, T=e^{im\phi})$ satisfy the commutation relations
 \begin{equation}
 \label{eq:qflatbis}
 \begin{gathered}
 i[j_m,j_n]=(m-n)j_{m+n},\\
 i[j_m,p_n]=(m-n)p_{m+n},\\
 i[p_m,p_n]=0.
 \end{gathered}
 \end{equation}

 The abstract ${\rm BMS}_3$ group \cite{Ashtekar:1996cd}\footnote{See
    e.g.~\cite{cantoni:1361}, \cite{Geroch:1971mz},
   \cite{1972RSPSA.330..517M}, \cite{mccarthy:1837} for similar
   considerations on the globally well-defined ${\rm BMS}_4$ group.} is
 the semi-direct product of the group of diffeomorphisms $ {\rm
   Diff}(S^1)$ of the circle with the abelian normal subgroup of tensor
 densities $\cF_1(S^1)$. If $\phi^\prime=f(\phi)$ denotes an element of the former
 and $a=\alpha(\phi)d\phi^{-1}$ an element of the latter, whose
 component transforms as
 $\alpha^\prime(\phi^\prime)=(\ddl{\phi^\prime}{\phi} \alpha)(\phi)$, the action
 of the diffeomorphisms on the tensor densities is given by
 \begin{equation}
 (f\cdot a)(\phi)=(\alpha\ddl{f}{\phi})(f^{-1}(\phi))d\phi^{-1}\label{eq:69}.
 \end{equation}
 The ${\rm BMS}_3$ group law is 
 \begin{equation}
   \label{eq:70}
   (f,a)\cdot (g,b)=(f\circ g, a+f\cdot b), 
 \end{equation}
 with a realization as coordinate transformations of $S^1\times
 \mathbb R$ of the form 
 \begin{equation}
   \label{eq:71}
   \phi^\prime=\phi^\prime(\phi), \quad
   u^\prime=\ddl{\phi^\prime}{\phi} (u + \alpha(\phi)),
 \end{equation}
 where now $\phi^\prime=f^{-1}(\phi)$. 

 \subsection{Gravitational results on the Minkowskian cylinder}
 \label{sec:results-cylinder-1}

 We summarize here results of \cite{Barnich:2010eb}. The BMS gauge
 consists in the metric ansatz
 \begin{equation}
   \label{eq:ojo}
    ds^2=e^{2\beta}\frac{V}{r} du^2-2e^{2\beta}
    dudr+r^2(d\phi-Udu)^2,
 \end{equation}
 for three arbitrary functions $\beta, V, U$.  In the flat case,
 assuming $\beta=o(1)=U$, the general solution to the equations of
 motion is
 \begin{equation}
   \label{eq:12}
   ds^2
 =\Theta(\phi)du^2-2dudr+2\Big[\Xi(\phi)+
 \frac{u}{2}\partial_{\phi}\Theta(\phi)\Big]du
   d\phi + r^2 d{\phi}^2. 
 \end{equation}
 The action of the asymptotic symmetries on solution space is given by 
 \begin{equation}
 \begin{split}
   -\delta\, \Theta & = Y\d_\phi \Theta+2 \d_\phi Y \Theta -
   2 \partial^3_\phi Y, \label{eq:B11}\\
   -\delta\, \Xi & = Y \partial_\phi\Xi+ 2 \d_\phi Y \Xi  +\half
   T\d_\phi\Theta+ \d_\phi T \Theta-\d^3_\phi
   T. 
 \end{split}
 \end{equation}
 The conserved surface charges computed at $\scri^+$ with respect to
 the null orbifold $\Theta=0=\Xi$ are
 \begin{equation}
   \label{eq:11}
 \begin{split}
 & Q_{T,Y}=\frac{1}{16 \pi G}
   \int_0^{2\pi}d\phi\, \big[T \Theta +
   2Y \Xi\big],\\
 & P^{\rm cycl}_m=\frac{1}{16 \pi G}
   \int_0^{2\pi}d\phi\, e^{im\phi} \Theta ,\quad
 J^{\rm cycl}_m=\frac{1}{8 \pi G}
   \int_0^{2\pi}d\phi\, e^{im\phi} \Xi,\\
 &  \Theta=8G\sum_m P^{\rm cycl}_m e^{-im\phi},\quad \Xi=4G\sum_m
 J^{\rm cycl}_m e^{-im\phi}.
 \end{split}
 \end{equation}
 The Dirac bracket charge algebra of the surface charge generators is
 then given by 
 \begin{equation}
 \label{eq:qflatter}
 \begin{split}
 & i\{J^{\rm cycl}_m,J^{\rm cycl}_n\}=(m-n)J^{\rm cycl}_{m+n},\\
 & i\{J^{\rm cycl}_m,P^{\rm cycl}_n\}=(m-n)P^{\rm
   cycl}_{m+n}+\frac{c_2}{12}m^3\delta^0_{m+n},\quad c_2=\frac{3}{G}\\
 & i\{P^{\rm cycl}_m,P^{\rm cycl}_n\}=0.
 \end{split}
 \end{equation}

 The cosmological solutions are characterized by $P^{\rm
   cycl}_m=\delta^0_m M\geq 0$ and $J^{\rm cycl}_m=\delta^0_m J\in
 \mathbb R$ while $P^{\rm cycl}_m=-\frac{c_2}{24}\delta^0_m, J^{\rm
   cycl}_m=0$ for Minkowski space-time.
 In particular, 
 \begin{equation}
   \label{eq:60bis}
   H=Q_{\partial_u}=P^{{\rm cyl}+}_0,\quad \cJ=Q_{\partial_\phi}=J^{{\rm cyl}+}_0,
 \end{equation}
 and thus, for the cosmological solutions, $H=M$, $\cJ=J$. 

 From the way they are constructed as surface integrals at $\scri^+$,
 the quantities $Q_{\partial_u},Q_{\partial_\phi}$ associated with the
 null vector $\partial_u$ and the space-like vector $\partial_\phi$ are
 respectively the Bondi mass and angular momentum, which are conserved
 in three dimensions due to the absence of news. As discussed before,
 in the particular case of the cosmological solutions, $\partial_u$ and
 $\partial_\phi$ are in addition Killing vectors.

 To recover more standard relations, it is useful to introduce the
 normalized variables $P_{++}(\phi)=-\frac{1}{8G} \Theta$,
 $J_{++}(\phi)=-\frac{1}{4G} \Xi$.

 \subsection{Mapping to the Euclidean plane}
 \label{sec:mapping-plane-1}

 Let $z=e^{i\phi}$. To go to the Euclidean plane, we also need $u=-i
 t_E$, but this is irrelevant for the charges which are time
 independent.  Infinitesimally, one has $z=\phi+ Y(\phi)$ and works to
 first order in $Y$, so that $Y=e^{i\phi}-\phi$ and $T=0$.

 Following the same computation as for the energy momentum tensor of a
 conformal field theory (see e.g.~Section 5.4.1 of
 \cite{DiFrancesco:1997nk}), the finite version of the first relation
 of \eqref{eq:B11} implies that $P_{++}$ transforms with the Schwarzian
 derivative,
 \begin{equation}
   \label{eq:19}
   P(z)=\left(\frac{dz}{d\phi}\right)^{-2}
   P_{++}(\phi)+\frac{c_2}{12}\{\phi;z\},\quad
 \{\phi;z\}=\frac{\frac{d^3\phi}{dz^3}}{\frac{d\phi}{dz}}
 -\frac{3}{2}\left(\frac{\frac{d^2\phi}{dz^2}}{\frac{d\phi}{dz}}\right)^2,
 \end{equation}
 so that
 \begin{equation}
   \label{eq:20}
    P_{++}(\phi)=-z^2 P(z)+\frac{c_2}{24}.
 \end{equation}
 The second equation of  \eqref{eq:B11} for $T=0$ then implies that
 there is no Schwarzian derivative term in the transformation of $J_{++}(\phi)$. 
 \begin{equation}
     \label{eq:21}
     J_{++}(\phi)=-z^2 J(z). 
 \end{equation}
 For $Y(\phi)$, because we are dealing with a vector field, we get
 $Y(\phi)=\epsilon(z)(iz)^{-1}$. The geometrical object is the tensor
 density $t=T(\phi)d\phi^{-1}$, so that $T(\phi)=\theta(z) (iz)^{-1}$, we
 then have 
 \begin{equation}
   \label{eq:22}
   \begin{split}
 & -\delta_{\epsilon,\theta} P=\epsilon \d P+2\d \epsilon P
 +\frac{c_2}{12} \d^3\epsilon,\\
 & -\delta_{\epsilon,\theta} J=\epsilon \d J+2\d \epsilon
 J+\theta\d P +2 \d\theta P+\frac{c_2}{12} \d^3 \theta,\\
 &Q_{\epsilon,\theta}=-\frac{1}{2\pi}\oint_{|z|=1} dz \Big[\theta(
 P-z^{-2}\frac{c_2}{24})+\epsilon J\Big],\\
 &P_m=\frac{1}{2\pi i} \oint_{|z|=1} dz\, z^{m+1} P,\quad J_m=
 \frac{1}{2\pi i} \oint_{|z|=1} dz\, z^{m+1} J ,\\
 &P(z)=\sum_{m}P_m z^{-m-2},\quad J(z)=\sum_{m} J_m z^{-m-2}\\
 &P_m^{\rm cyl}=P_m-\frac{c_2}{24}\delta^0_m, \quad J_m^{\rm
   cyl}=J_m. 
   \end{split}
 \end{equation}
 In terms of $P_m,J_m$, the algebra is as in \eqref{eq:qflatter} with
 the central term changed from $\frac{c_2}{12}m^3\delta^0_{m+n}$
 to $\frac{c_2}{12}m(m^2-1)\delta^0_{m+n}$. 

 Minkowsi space-time corresponds to $P_m=0=J_m$. The assumption is now
 that this solution corresponds to the vacuum state, then there is a
 mass gap and the cosmological solutions correspond to the other
 relevant states.  In terms of $P_0$, the vacuum state is at zero
 eigenvalue and then the other relevant states have eigenvalues
 greater or equal to $\frac{c_2}{24}$.

 \subsection{Cardy-like formula for the flat-space partition
   function}
 \label{sec:cardy-like-formula}

 Consider\footnote{The considerations of this section have been
   elaborated on the basis of an argument by S. Detournay, T. Hartman
   and D. Hofman which has appeared in \cite{Detournay:2012pc} after
   the current work has been accepted for publication.} the partition
 function on the torus defined as
 \begin{equation}
   \label{eq:61}
   Z(\beta,\mu)={\rm Tr}_{\cH} e^{-\beta (H+\mu \cJ)}. 
 \end{equation}

 By introducing a temperature $\beta$, one introduces a length scale
 into the system, which can be taken to be $l$, so that $\tilde
 \beta=\frac{\beta}{l}$ is dimensionless. At this stage, $l$ has nothing
 to do with a cosmological radius. Consider then the
 complex plane with $z=\phi+i\frac{u}{l}$ and the cylinder defined by
 the periods $\omega_1,\omega_2\in \mathbb C$. The
 ${\rm BMS}_3$ transformation $\phi^\prime=\frac{1}{\omega_1}\phi$,
 $\alpha(\phi)=0$ implies $z^\prime=\frac{1}{\omega_1} z$ so that one can
 set $\omega_1$ to $1$ and work in terms of the modular parameter
 $\tau=\frac{\omega_2}{\omega_1}$. The question is then whether a
   $PSL(2,\mathbb Z)$ transformation $\left(\begin{smallmatrix} a&b\\
       c&d\end{smallmatrix}\right)$ of the torus can be induced from a ${\rm
   BMS}_3$ transformation, or in other words, whether 
 \begin{equation}
 \phi^\prime+i\frac{u^\prime}{l}=\frac{az +b}{cz+d}=
 \frac{(a\phi+b)(c\phi+d)+l^{-2} u^2
   ac+i\frac{u}{l}}{(c\phi+d)^2+l^{-2} u^2 c^2}.
 \label{eq:13a}
 \end{equation}
 When $c=0$, this is possible by choosing $\phi^\prime=\phi$,
 $\alpha(\phi)=il(\phi-a(a\phi+b))$. When $c\neq 0$, this is possible only up
 to terms of order $l^{-2}$, in which case
 $\phi^\prime=\frac{a\phi+b}{c\phi+d}$, $\alpha(\phi)=0$. 

 It follows from section \bref{sec:euclidean-solution} that the modular
 parameter relevant for the cosmological solution is
 $\tau=\frac{1}{2\pi}(\Phi_E+\frac{i\beta}{l})=\frac{\beta}{2\pi}(\mu_E+\frac{i}{l})$
 with $\mu=-i\mu_E$. Invariance under the modular transformation
 $\tau\to -\frac{1}{\tau}$ would imply that
 \begin{equation}
   \label{eq:85}
   Z(\beta,\mu_E)=Z(\frac{4\pi^2}{\beta(l^{-2}+\mu^2_E)},
   -\frac{4\pi^2\mu_E}{\beta^2(l^{-2}+\mu^2_E)}). 
 \end{equation}
 In terms of $\Phi_{E}=\beta\mu_{E}$, this invariance takes the
 form 
 \begin{equation}
   \label{eq:5}
   \beta\to \frac{4\pi^2}{\beta(\frac{\Phi^2_E}{\beta^2}+l^{-2})},
   \quad \Phi_E\to
   -\frac{4\pi^2}{\Phi_E(1+l^{-2}\frac{\beta^2}{\Phi^2_E})}. 
 \end{equation}
 It thus follows that the partition
 function of a ${\rm BMS}_3$-invariant theory is expected to satisfy
 \begin{equation}
   \label{eq:16}
   Z(\beta,\mu_E)=Z(\frac{4\pi^2}{\beta\mu^2_E},
   -\frac{4\pi^2}{\beta^2\mu_E}),
 \end{equation}
 or in terms of $\Phi_E$,
 \begin{equation}
   \label{eq:10}
   \beta\to \frac{4\pi^2\beta}{\Phi^2_E},\quad \Phi_E\to
   -\frac{4\pi^2}{\Phi_E}. 
 \end{equation}
 Taking into account the mass gap, one finds in the high-temperature limit $\beta\to 0$,
 \begin{equation}
   \label{eq:18}
   \ln Z_{\rm Cardy}(\beta,\mu_E)=\frac{4\pi^2}{\beta\mu^2_E}\frac{c_2}{24}=\frac{\pi^2}{2G\beta
     \mu_E^2 },
 \end{equation}
 which agrees with \eqref{eq:14}.

 \section{Flat-space limit of ${\rm AdS}_3$ results}
 \label{sec:flat-space-limit}

 \subsection{Symmetries and charges}
 \label{sec:symmetries-charges}

 In order to compare flat-space and ${\rm AdS}_3$ results
 \cite{Barnich:2012aw}, it is useful to present both in the same ${\rm
   BMS}$ gauge rather than using the more usual Fefferman-Graham gauge
 for the latter. The correct scaling of space-time coordinates that
 gives the limit then turns out to be a modified Penrose limit. On the
 level of the algebra of symmetries and charges, this approach shows in
 detail and in spacetime terms how the contraction, identified
 previously on purely algebraic grounds in \cite{Barnich:2006avcorr},
 comes about: in a first step, the two copies of the Virasoro algebra
 $L^\pm_m$ with equal central charges $c^\pm=\frac{3l}{2G}=c$ in the
 gravitational case, is presented in terms of the redefined generators
 \begin{eqnarray}
   \label{generators}
  P_{m}=\frac{1}{l}(L^{+}_{m}+L^{-}_{-m}),  \: \: J_{m}=L^{+}_{m}-L^{-}_{-m},
  \end{eqnarray}
 and reads
 \begin{equation}
 \label{eq:qflat}
 \begin{gathered}
 i\{J_m,J_n\}=(m-n)J_{m+n}+\frac{c^+-c^-}{12}m(m^2-1)\delta^0_{m+n},\\
 i\{J_m,P_n\}=(m-n)P_{m+n}+\frac{c^++c^-}{12\ell}m(m^2-1)\delta^0_{m+n},\\
 i\{P_m,P_n\}=\frac{1}{l^2}\big((m-n)J_{m+n}+
 \frac{c^+-c^-}{12}m(m^2-1)\delta^0_{m+n}\big),
 \end{gathered}
 \end{equation}
 In the second step, at fixed generators, the limit $l\to \infty$ is
 taken and reduces to the flat-space result.

 When normalized with respect to the $M=0=J$ BTZ black hole, the
 Hamiltonian is $H=\frac{1}{l}(L_0+\bar L_0-\frac{c}{12})$. The mass
 gap with ${\rm AdS}_3$ spacetime is the same than the one in flat
 space between the cosmological solutions and Minkowski spacetime and
 given by $\frac{c}{12 l}=\frac{c_2}{24}=\frac{1}{8G}$, independently
 of $l$.

 \subsection{Thermodynamics}
 \label{sec:btz-thermodynamics}

 For the BTZ black holes, the standard ADM form is
 \begin{equation}
   \label{eq:26a}
 \begin{gathered}
   ds^2=-N^2dt^2+N^{-2}dr^2+r^2(d\varphi+N^\varphi dt)^2,\\ 
 N^2=\frac{r^2}{l^2}-8MG +\frac{16 G^2 J^2}{r^2},\quad
 N^\varphi=\frac{4GJ}{r^2}.
 \end{gathered}
 \end{equation}
 Defining 
 \begin{equation}
   \label{eq:78}
   r^2_{\pm}=4GMl^2\Big[1\pm\sqrt{1-\frac{J^2}{M^2l^2}}\Big],\quad
   M=\frac{r_+^2+r^2_-}{8Gl^2},\quad J=\frac{r_+r_-}{4Gl},
 \end{equation}
 temperature, angular velocity and Bekenstein-Hawking entropy are given
 by 
 \begin{equation}
   \label{eq:77}
 \begin{split}
  & T_H=\frac{1}{\beta}=\frac{r^2_+-r^2_-}{2\pi l^2 r_+},\quad
   \mu=\Omega_H=-\frac{r_-}{r_+ l},\quad S_{BH}=\frac{2\pi r_+}{4G},
 \end{split} 
 \end{equation}
 with a first law of the form 
 \begin{equation}
 dM=T_HdS_{BH}-\Omega_H dJ\label{eq:7}.
 \end{equation}
 The Bekenstein-Hawking entropy, and thus also the matching Cardy
 formulas for the entropy \eqref{eq:84} below, can obviously not be
 obtained as the limit $l\to\infty$ of the ${\rm AdS}$ case since
 $r_+(M,J,G;l)$ is pushed out to infinity and does not have a good
 limit.

 Left and right temperatures are defined through 
 \begin{equation}
   \label{eq:80}
   T_+= \frac{T_H}{1+l\Omega_H}=\frac{r_++r_-}{2\pi l^2}, \quad
   T_-=\frac{T_H}{1-l\Omega_H}=\frac{r_+-r_-}{2\pi l^2}. 
 \end{equation}
 Inverting the relations in terms of inverse temperature and chemical
 potential, one gets
 \begin{equation}
   \label{eq:79}
  r_+=\frac{2\pi }{\beta(l^{-2}-\mu^2)},\quad r_-=-\frac{2\pi
    l\mu}{\beta(l^{-2}-\mu^2 )}, \quad S_{BH}(\beta,\mu)=\frac{\pi^2
    }{G\beta(l^{-2}-\mu^2 )}. 
 \end{equation}
 In this case, due to the standard form of the first law, the
 Massieu function for the Bekenstein-Hawking entropy is 
 \begin{multline}
   \label{eq:83}
 \ln  Z_{BH}(\beta,\Phi)=S_{BH}-\beta M-\Phi J, \\ \ln  Z_{BH}=\frac{\pi^2\beta
    }{2G(l^{-2}\beta^2-\Phi^2 )}=\frac{\pi^2
    }{2G\beta (l^{-2}-\mu^2 )}. 
 \end{multline}

 Its flat space-limit agrees with the one for the cosmological solution
 \eqref{eq:14}. 

 The thermodynamics of the inner (Cauchy) horizon of the BTZ
 black holes has also been discussed recently 
 \cite{Detournay:2012ug} and is given by 
 \begin{equation}
   \label{eq:24}
   T_H^-=\frac{1}{\beta^-}=\frac{r^2_+-r^2_-}{2\pi l^2 r_-}, \quad  
 \mu^-=\Omega^-_H=-\frac{r_+}{r_- l},\quad S^-=\frac{2\pi r_-}{4G},
 \end{equation}
 with a first law of the form 
 \begin{equation}
   dM=-T^-_HdS^- -\Omega^-_H dJ\label{eq:7a}. 
 \end{equation}
 As a side remark, note that when introducing $T_H^+=T_H$, we have
 $\frac{1}{T_H^\pm}=\half(\frac{1}{T_-}\pm \frac{1}{T_+})$. 
 Inverting gives in this case
 \begin{multline}
   \label{eq:27}
    r_-=\frac{2\pi }{\beta^-(({\mu^-})^2 -l^{-2})},\ r_+=-\frac{2\pi
    l\mu^-}{\beta^-(({\mu^-})^2 -l^{-2})}, \\ S^-(\beta^-,\mu^-)=\frac{\pi^2
    }{G\beta^-(({\mu^-})^2 -l^{-2})}. 
 \end{multline}
 Again, taking into account the unusual form of the first law, the 
 Massieu function is 
 \begin{multline}
   \label{eq:14a}
   \ln  Z^-(\beta^-,\Phi^-)=-S^--\beta^- M-\Phi^- J , \\ \ln
   Z^-=\frac{\pi^2\beta^-}{2G(l^{-2}(\beta^-)^2-({\Phi^-})^2)}=
 \frac{\pi^2}{2G\beta^-(l^{-2}-({\mu^-})^2)}. 
 \end{multline}
 If one chooses the negative sign for the Hawking temperature,
 $T_H^-=\frac{1}{\beta^-}=\frac{r^2_--r^2_+}{2\pi l^2 r_-}$, the signs
 in the first law and the Legendre transform become standard,
 $r_+(\beta_-,\mu_-)$, $r_-(\beta_-,\mu_-)$ change sign, while the
 final expression for $\ln Z^-$ is unchanged.

 In all cases, the Massieu function is universal in the sense that it
 does not depend on whether one derives it from the Bekenstein-Hawking
 entropy of the inner or the outer BTZ horizon.

 The horizon of the cosmological solution is the limit of the inner
 horizon of the BTZ black hole,
 \begin{equation}
   \label{eq:92}
   r_C=\lim_{l\to \infty}r_-(M,J,G;l). 
 \end{equation}
 Furthermore, all thermodynamic variables of the cosmological solutions
 are precisely the flat-space limit $l\to \infty$ of the variables of
 the inner horizon of the BTZ black hole.

 \subsection{Cardy-like formula in the grand canonical ensemble}
 \label{sec:cardy-formula}

 As discussed in
 \cite{Strominger:1998eq,Maldacena:1998bw,Carlip:1998qw}, when using
 the Cardy formulas
 \begin{equation}
   \label{eq:84}
   S_{\rm Cardy}=\pi\sqrt{\frac{2c^+ L_0^{{\rm cycl}
         +}}{3}}+\pi\sqrt{\frac{2c^- L_0^{{\rm cycl} -}}{3}}=\frac{\pi^2l}{3}(c^+T_+ +
  c^-T_-), 
 \end{equation}
 one gets agreement, 
 \begin{equation}
   \label{eq:82}
   S_{\rm Cardy}=S_{BH}. 
 \end{equation}
 As a side remark, we also notice that
 $S_{BH}^-=\frac{\pi^2l}{3}(c^+T_+ - c^-T_-)$.

 Instead of a Cardy formula for the entropy, one can derive an
 equivalent formula for the partition in the standard way. Indeed, the
 partition function can be written as
 \begin{equation}
   \label{eq:61a}
   Z[\beta,\mu]={\rm Tr}_{\cH} e^{-\beta (H+\mu \cJ)}={\rm Tr}_{\cH}
   q^{L_0-\frac{c}{24}} \bar q^{\bar L_0-\frac{c}{24}}=Z[\tau,\bar\tau],
 \end{equation}
 where $q=e^{2\pi i\tau}$ and
 \begin{equation}
   \label{eq:62}
   \tau=\frac{\beta}{2\pi}(\mu_E+\frac{i}{l}),
 \end{equation}
 with 
 \begin{equation}
   \label{eq:15}
    \mu=-i\mu_E. 
 \end{equation}
 Modular invariance now implies \eqref{eq:85} for all values of $l$,
 the flat-space limit being \eqref{eq:16}. 

 In the high temperature limit $\beta\to 0$ one then finds  
 \begin{equation}
   \label{eq:86}
  \ln  Z_{\rm Cardy}(\beta,\mu_E)=
  \frac{4\pi^2}{\beta(l^{-2}+\mu_E^2)}\frac{c}{12 l},  
 \end{equation}
 which agrees with \eqref{eq:83} when using \eqref{eq:15}. In addition,
 its flat-space limit gives \eqref{eq:18}, as it should. 

 \section{Discussion}
 \label{sec:discussion}

 In the context of non-relativistic versions of the AdS/CFT
 correspondence in three dimensions, the two dimensional Galilean
 conformal algebra $\mathfrak{gca}_2$ algebra plays a prominent role
 \cite{Bagchi:2009ca,Bagchi:2009my,Bagchi:2009pe}.  Based on
 \cite{Barnich:2006avcorr} and the holographic interpretation of the
 flat space asymptotic structure in \cite{Barnich:2010eb} in terms of
 the first two representations discussed in section
 \bref{sec:asympt-flat-3d}, it has been pointed out in
 \cite{PhysRevLett.105.171601} that the $\mathfrak{bms}_3$ algebra is
 isomorphic to $\mathfrak{gca}_2$. Referring to this kind of symmetry
 based flat space holography as a ${\rm BMS/GCA}$ correspondence is
 thus misleading. This is so not only from a chronological but also
 from a physical point of view. Indeed, contrary to what the wording of
 \cite{Bagchi:2012cy} might suggest, the correct scaling of coordinates
 that implements the algebraic contraction \cite{Barnich:2006avcorr} of
 the two copies of the Virasoro algebras to the $\mathfrak{bms}_3$
 algebra in either bulk or boundary space-time terms, has nothing to do
 with a non-relativistic limit but rather, as shown through a detailed
 bulk analysis in \cite{Barnich:2012aw}, with a modified Penrose limit.

 Of course, the isomorphism of algebras means that group theoretic
 results on $\mathfrak{gca}_2$ are very relevant for flat-space
 holography. Other results may be transposed as well. For instance, the
 remnant of modular invariance and the resulting Cardy-like formula for
 the partition function that comes from the non-relativistic
 contraction discussed in \cite{Hotta:2010qi} corresponds to exchanging
 the role of $\beta$ and $\Phi_E$ followed by changing the signs in
 \eqref{eq:10}. This is consistent with the different roles played by
 $M_0,L_0$ and $P_0,J_0$ in both contexts.

 Apart from the interest for asymptotically flat three-dimensional
 gravity and holography in this context, one of the more intriguing
 points of the analysis is the universality of the Massieu function
 with respect to the inner and outer BTZ horizons. The natural question
 that arises is whether this universality holds in more general cases
 and for other horizons as well. It is straightforward to check
 \cite{barnich2102:zz} that it holds also for instance for the BTZ
 black hole in topological massive gravity or, with a bit more work,
 for the Kerr black hole, for which the thermodynamics at the inner
 horizon was originally studied in
 \cite{springerlink:10.1007/BF02743435,springerlink:10.1007/BF02739031}. The
 results on the universality of the form of the first law at the inner
 horizon should thus really be understood as the proof, for these
 cases, of the universality of the Massieu function. What this implies
 at the quantum level, maybe not quite so surprisingly, is that it is
 really the partition function that is universal.

 \section*{Acknowledgements}
 \label{sec:acknowledgements}

 \addcontentsline{toc}{section}{Acknowledgments}

 The author is grateful to S. Detournay for an illuminating
 discussion. He thanks R.~Argurio, M.~Ba\~nados, A.~Gomberoff,
 H.~Gonz\'alez, P.-H.~Lambert, B.~Oblak, D.~Tempo, C.~Troessaert and
 R.~Troncoso for extensive collaborations on and discussions of
 relevant background material. This work is supported in part by the
 Fund for Scientific Research-FNRS (Belgium), by the Belgian Federal
 Science Policy Office through the Interuniversity Attraction Pole
 P6/11, by IISN-Belgium, by ``Communaut\'e fran\c caise de Belgique -
 Actions de Recherche Concert\'ees'' and by Fondecyt Projects
 No.~1085322 and No.~1090753.


\def\cprime{$'$}
\providecommand{\href}[2]{#2}\begingroup\raggedright\endgroup

\end{document}